\def\year{2021}\relax
\documentclass[letterpaper]{article} 
\usepackage{aaai21}  
\usepackage{times}  
\usepackage{helvet} 
\usepackage{courier}  
\usepackage[hyphens]{url}  
\usepackage{graphicx} 
\usepackage{booktabs}
\usepackage{amsmath}
\usepackage{natbib}  
\usepackage{caption} 
\usepackage{url}
\makeatletter
\g@addto@macro{\UrlBreaks}{\UrlOrds}
\makeatother
\usepackage{comment}
\usepackage{balance}       
\usepackage{graphics}      

\title{User Acceptance of Gender Stereotypes in Automated Career Recommendations}
\author {
    Clarice Wang\textsuperscript{\rm 1},
    Kathryn Wang\textsuperscript{\rm 2},
    Andrew Bian\textsuperscript{\rm 3},
    Rashidul Islam\textsuperscript{\rm 4},
    Kamrun Naher Keya\textsuperscript{\rm 4},
    James Foulds\textsuperscript{\rm 4},
    Shimei Pan\textsuperscript{\rm 4} \\
}
\affiliations {
    \textsuperscript{\rm 1} The Harker School, 
    San Jose, CA \\
    \textsuperscript{\rm 2} The Dalton School,
    New York, NY \\
    \textsuperscript{\rm 3} River Hill High School,
    Clarksville, MD \\
    \textsuperscript{\rm 4} University of Maryland, Baltimore County, 
    Baltimore, MD \\
    22claricew@students.harker.org, wang.kathryn.wang@gmail.com, abian1960@inst.hcpss.org,\\
    \{islam.rashidul, kkeya1, jfoulds, shimei\}@umbc.edu \\
}

\begin{document}

\maketitle

\begin{abstract}
Currently, there is a surge of interest in {\it fair Artificial Intelligence (AI) and Machine Learning (ML)} research which aims to mitigate discriminatory bias in AI algorithms, e.g. along lines of gender, age, and race. While most research in this domain focuses on developing fair AI algorithms, in this work, we show that a fair AI algorithm on its own may be insufficient to achieve its intended results in the real world. Using career recommendation as a case study, we build a fair AI career recommender by employing  gender debiasing machine learning techniques.  Our offline evaluation showed that the debiased recommender makes fairer career recommendations without sacrificing its accuracy.  Nevertheless, an online user study of more than 200 college students revealed that participants on average prefer the original biased system over the debiased system. Specifically, we found that perceived gender disparity is a determining factor for the acceptance of a recommendation. In other words, our results demonstrate we cannot fully address the gender bias issue in AI recommendations without addressing the gender bias in humans. 

\end{abstract}

\section{Introduction}
Artificial Intelligence (AI) is increasingly used in consequential decision making, but recent discoveries have shown that AI systems 
often exhibit discriminatory bias in their behavior, 
particularly along gender, age, and racial lines~\cite{dastin2018amazon,noble2018algorithms, angwin2016machine}. 
For example, an AI tool that helps judges assess the risk of an incarcerated individual committing a crime in the future was found to be  biased against African Americans~\cite{angwin2016machine}. In other domains including personalized search, ads and recommendation, AI systems lead to skewed outcomes~\cite{ali2019discrimination}. 
AI models trained on text data have been found to encode gender stereotypes such as associating computer programming with men and homemaking with women~\cite{bolukbasi2016man}, which could potentially impact  automated career counseling and hiring decisions. Indeed,
Amazon had to scrap its AI recruiting tool because it was found to be biased against women~\cite{dastin2018amazon}.  





Since biased AI systems can be discriminatory against vulnerable populations in our society and/or reinforce harmful stereotypes, 
there are strong motivations to develop AI debiasing interventions~\cite{campolo2017ai}. We 
argue that there are two aspects which must be considered when it comes to mitigating AI bias: the algorithmic aspect and the human-AI interaction aspect. 
Most existing approaches in the AI community focus on the algorithmic aspect. The mission of the field has primarily been to develop \emph{(a)} new metrics that quantify fairness~\cite{dwork2012fairness,Kusner2017}, and \emph{(b)} new machine learning techniques that remove bias~\cite{zhang2018mitigating,dev2019attenuating,foulds2020bayesian}. On the other hand, the human-fair AI interaction aspect as well as its broader social context is equally important and significantly understudied. 
For example, in many contexts such as targeting ads on search and social network platforms it is well understood that there is a tension between building a fair system and achieving the platform's own revenue goals~\cite{Miller2019}, and this tension cannot be resolved by algorithms alone. 
As bias in AI may arise from the human side of the socio-technical system via systemic bias and/or human prejudice encoded in data \cite{barocas2016big}, will bias mitigation be effective if we focus only on removing the bias in AI without addressing the bias in humans?

%

In this research, we systematically study the interplay between AI debiasing techniques and humans, using AI career recommendation as a case study. Career choice is a major part of human life, as it often defines one's economic success, social standing, and quality of life. Humans have a tendency to associate masculine and feminine traits with specific careers, which results in the perception that certain genders are better suited for certain occupations~\cite{white2006}.  Machines that learn from career data are thus expected to be influenced by the resulting gender gaps in career choices unless bias mitigation techniques are performed~\cite{dastin2018amazon}. 

We first dive into the algorithmic aspect of the problem, using machine learning to systematically mitigate bias in AI systems so that they do not reinforce harmful stereotypes.  Second, we examine the human-fair AI interaction aspect of the problem.  
We perform a user study to investigate whether users will typically prefer a fair AI system over a biased one. Users may have their own biases, and our results show that users' acceptance of a debiased AI system can be influenced by their own biases. 
%
%
%
%
%
%
The following are the main contributions of the paper.
\begin{itemize}
    \item  To the best of our knowledge, this is the first study on how human bias interferes with the effectiveness of a fully implemented fair AI career. 
   \item As a case study, we develop a debiased recommender system which mitigates gender bias in career recommendations~\cite{Anonymous1}. Our offline evaluation 
   shows that the debiased recommender is fairer than the gender-aware (biased) recommender without any loss of prediction accuracy. 
    \item We conduct an online user study with over 200 participants (college students) to understand their acceptance of the debiased system. The results indicate that participants in general prefer the gender-aware (biased) career recommender over the gender-debiased one. We analyzed the role a participant's own bias plays in his or her  acceptance of the recommendations. Our results indicate that the perceived gender disparity in a recommended career is significantly correlated with its acceptance.
\end{itemize}
In the rest of the paper, we describe related literature, the implementation of a debiased machine learning algorithm for fair career recommendation, an offline evaluation of the system,  an online user study for understanding the relationship between human bias and the  acceptance of a fair AI system. 

\section{Related Work}
In this section, we  summarize  recent work on fair AI and ML, review  social science studies on the relationship between gender bias and career decisions, and briefly discuss the work in the Human-Computer Interaction (HCI) community on AI bias/fairness. 

\subsection{Fair AI and ML Research}
Recently, there has been a sharp focus in the AI community on how to prevent AI from perpetuating or, worse, exacerbating social unfairness.  Most efforts concentrate on (1) developing metrics to quantify the bias in data as well as in ML algorithms, and (2) developing fair ML algorithms that mitigate these biases. 

It is difficult to develop a universal definition of fairness because fairness/bias is a complex,  multifaceted concept whose definition heavily depends on the social, culture and application context. Consequently, many definitions have been proposed. In fact, AIF360, the IBM open source platform for fair machine learning, has several dozen metrics for fairness/bias~\cite{Bellamy19}. Among them, some are about {\it individual fairness} and others are about {\it group fairness}. 

Individual fairness seeks to ensure similar individuals get similar outcomes. Widely used individual fairness measures include {\it Fairness Through Awareness} ("An algorithm is fair if it gives similar predictions to
similar individuals")~\cite{dwork2012fairness}, {\it Fairness Through Unawareness} ("An algorithm is fair as long as any protected attributes such as race, age, gender are not explicitly used in the decision-making process")~\cite{gajane2017formalizing}, and {\it Counterfactual Fairness} ("a decision is fair towards an individual if it is the same in both the actual world and a counterfactual world where the individual belonged to a different demographic group")~\cite{Kusner2017}. 

Group fairness partitions a population into
groups defined by protected attributes (or intersections of protected attributes) and seeks to ensure that statistical measures of outcomes are equal across groups/subgroups. Widely used group fairness measures include {\it Demographic Parity} (``The likelihood of a positive outcome should be the same regardless of whether the person is in the protected group'')~\cite{dwork2012fairness}, {\it Equalized Odds} (``the probability of a person in the positive class being correctly assigned a positive outcome and the probability of a person in a negative class being incorrectly assigned a positive outcome should both be the same for the protected and unprotected group members'')~\cite{hardt2016equality}, {\it Equal Opportunity} (``the probability of a person in a positive class being assigned to a positive outcome should be equal for both protected and unprotected group members'')~\cite{hardt2016equality} and {\it Differential  Fairness} (``the probabilities of the outcomes will be similar with respect to different subgroups defined by the intersections of multiple protected attributes such as race, gender and race'')~\cite{foulds2020intersectional}. 
Most fairness measures are defined for classification tasks. There are fairness measures for other tasks. For example, {\it Non-parity Unfairness} ~\cite{Yao2017FairCF} is designed to evaluate the fairness of recommender systems.


In terms of mitigating bias in AI systems, one focus is to remove the bias from the data that is used to train the AI models. For example, vector projection-based bias attenuation method is used to remove bias from word embeddings~\cite{dev2019attenuating}. Since word embeddings are widely used as features to train downstream  natural language processing models, debiased word embedding improves the fairness of the applications. There is also a large body of work that optimizes ML models under both traditional accuracy-based and new fairness-based objectives~\cite{woodworth2017learning, zafar2017fairness,agarwal2018reductions,foulds2020bayesian}. Furthermore, adversarial learning is used to improve model accuracy and at the same time minimize an adversary's chance of finding out protected attributes (e.g., gender and race)~\cite{zhang2018mitigating}.

\subsection{Social Science Research on Gender Bias and Careers}
According to Glick et al~\cite{glick1999gender}, gender, or the cultural construction of sex differences, is  the ``most automatic, pervasive and earliest learned'' categorization that shapes social relations and identities. Social science research on gender bias and stereotypes in career choices consistently finds that gender-based differences in career selection exist even when controlling for measured competency and ability~\cite{correll2001gender}. Career-related gender bias exists across country, culture and age. For example,  even as kindergartners, girls select mostly traditional female careers such as teaching and nursing ~\cite{stroeher1994sixteen}.  Scottish pupils were found to perceive Truck Driver, Engineer, Plumber/Electrician, Laborer, Armed Forces as "male" jobs while Nurse and Care Assistant "female" jobs. Boys, but especially girls, have strong preferences against working in sectors and industries that are traditionally the domain of the opposite sex~\cite{mcquaid2004gender}. 

Many theories have been developed to explain why  gender bias exists in career selection. Some of them focus on psychological constructs (i.e., variables at the level of individuals), while others focus on socioeconomic conditions and cultural understandings of gender roles.

{\it Social Cognitive Theory}~\cite{bandura1977self} and {\it Social Cognitive Career Theory}~\cite{lent1994toward} are the most influential social cognitive frameworks for understanding individual human behavior as well as career decisions. They posit that human behaviour is primarily explained through self-efficacy beliefs, outcome expectations, and goal representations. Self-efficacy beliefs refer to ``people’s judgment of their capabilities to organize and execute courses of action required to attain designated types of performances.'' Outcome expectations concern a ``person’s estimate that a given behaviour will lead to a certain outcome.''  Goal representations are defined as ``determinations of individuals to engage in a particular activity.''  Of the three determinants, self-efficacy 
has the strongest influence on behavior.  Gender difference in self-efficacy beliefs may explain observed gender bias in career choice. For example, it was found that women possess lower levels of mathematics confidence than men because women had fewer learning possibilities and role models to stimulate them~\cite{bandura1978reflections,lent1994toward}.

In addition to psychological constructs, social and cultural beliefs about gender may also influence  career choice of men and women. For example, gender beliefs are cultural schemas for interpreting or making sense of the social world. They represent what we think ``most people'' believe or accept as true about the categories of ``men'' and ``women.'' Substantial evidence indicates that certain careers (e.g., mathematics) are often stereotyped as “masculine”~\cite{meece1982sex,hyde1990gender}.  This cultural belief about gender channels men and women in substantially different career directions since it impacts the self-efficacy of individuals (e.g., belief about their own mathematical competence)~\cite{correll2001gender}.

In terms of overcoming gender bias in career decisions, the most commonly cited interventions include the availability of role models in the same social circle, especially same sex role models for women~\cite{lockwood2006someone,hill2014career,mcquaid2004gender}. Encouragement from friends and family~\cite{leaper2019helping} also improves self-efficacy.

\subsection{HCI Research on AI Fairness/Bias}
Recent work on AI fairness/bias in the HCI community mostly focuses on identifying and analyzing  biases in  existing AI systems such as image search~\cite{otterbacher2017competent,kay2015unequal}, social media analysis~\cite{johnson2017effect}, image and persona generation~\cite{salminen2019detecting,salminen2020analyzing}, sentiment analysis~\cite{diaz2018addressing}, text mining~\cite{cryan2020detecting} and natural language generation~\cite{strengers2020adhering}. 

There are also some work on fairness perception and definition. Wang et al~\cite{wang2020factors} and McDonald et al~\cite{McDonald2020} conducted studies to better understand the perception of fairness. Hou et al~\cite{hou2017factors} and Woodruff et al~\cite{woodruff2018qualitative} explored how intended users, especially those marginalized by race or class, feel about algorithmic fairness. Dodge et al~\cite{Dodge2019} conducted an empirical study on how people judge the fairness of ML systems and how explanations impact that judgment.  Chen et al~\cite{chen2020general} tried to quantify fairness/biases by measuring the difference of its data distribution with a reference dataset using Maximum Mean Discrepancy.

There is a rich body of work on mitigating bias in AI systems through the use of better system designs. One principled approach is known as equitable and inclusive co-design, which is about engaging diverse stakeholders especially underrepresented minorities directly in the design process~\cite{madaio2020co,skinner2020children,vorvoreanu2019gender,metaxa2018gender}. Furthermore, better tooling~\cite{cramer2019translation,yan2020silva} and better algorithms~\cite{strengers2020adhering,barbosa2019rehumanized}  also help address the problem.

So far, there is little work focusing on what happens {\it after} biases in AI systems are identified and systematically removed. Does it automatically achieve its intended societal impact? Our work explores this domain.

\begin{figure*}[t]
\centering
\includegraphics[scale=.27]{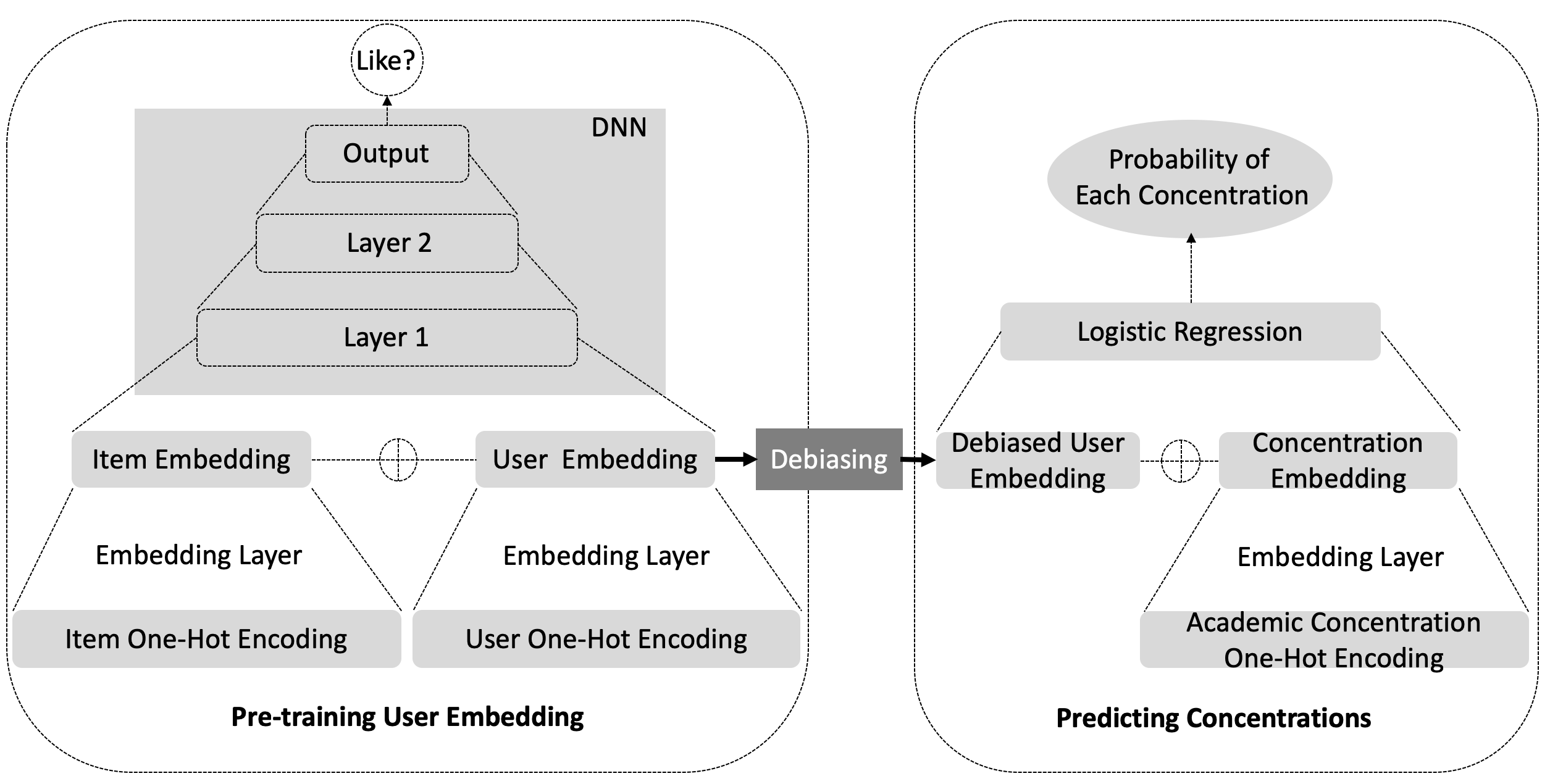}
\caption{The Architecture of a Gender-Debiased Career Recommender}
\label{fig:arch}
\end{figure*}

\section{Debiasing Career Recommendations}
We use an AI career recommendation as a case study to illustrate how an AI recommender which uses  state-of-the-art debiasing techniques to systematically remove gender stereotypes from its recommendation, may not produce intended outcomes.  In this study, we focus on college major recommendation because the current subjects of our study were college students, to whom the choice of a major is one of the most consequential career decisions they make. 

Recommender systems have gained widespread acceptance in the era of the internet, social network, and e-commerce. The basic idea of recommender systems is to infer user interest based on user-generated data. For example,  collaborative filtering, a key method used in recommender systems, is based on the assumption that similar users have similar preferences of items~\cite{sarwar2001item}. Thus, an e-commerce recommender will recommend a product to a customer if it was purchased by customers who gave similar ratings to other products in the past. Following the same idea, an AI career recommender suggests similar college majors to people who share similar interests.    


A major source of bias/unfairness in machine learning outcomes arises from biases in the data~\cite{barocas2016big,mehrabi2019survey,suresh2019framework}. A machine learning model trained on biased data may lead to unfair/biased predictions. For example, user preference data may encode real-world human biases (e.g., gender or racial biases). As a result, the system may inherit the bias into its recommendations. 
In the following, we describe the methodology used to implement a debiased algorithm for fair college major recommendation. 

\subsection{Backend Algorithm Implementation}


The system is designed to make career recommendations based on an individual's interests. The input to the system is a user's interests indicated on Facebook. We chose  the Facebook data because it contains a wide range of items a person can like such as books, hobbies, thoughts, movies, music and sports. The output contains the top college major recommended by our system such as computer science, biology and psychology. We employed an existing Facebook dataset widely used in social media research~\cite{matz2017psychological,youyou2015computer,schwartz2013personality,kosinski2015facebook}.
 
Since the academic concentrations in our data are declared by Facebook users, they are quite noisy (e.g., ``Defense Against the Dark Arts'' is a declared academic concentration). To prepare the Facebook data to train our  recommender, we filtered out academic concentrations that occurred less then 3 times in the data. The final data used in training and testing the backend system contains a total of 16,619 users (of which 60\% are female, 40\% are male and no gender non-binary individuals), 1,380 unique concentrations, 140K+ unique items that a user can like and 3.5 million+ user-interest pairs.

To develop a college major recommender, first, we train a neural collaborative filtering (NCF)~\cite{he2017neural} model for predicting the items a user ``likes,'' encoded as 1, or 0 if otherwise.  A user's gender is not taken into account during the training of the NCF model.  We also included $10\%$ negative instances from those user-interest pairs marked as ``0.''  In the input layer, the users and items are one hot-encoded.  They are mapped into two separate embedding layers with embedding size of 100 (user and item embedding). Since NCF adopts two pathways to model users and items, the user and item embeddings are combined by concatenation. One hidden layer with 10 linear units is added on the concatenated vector along with dropout regularization of probability 0.1, followed by a linear output layer. Finally, we train the model by optimizing MSE loss using \emph{Adam} in batch mode with a learning rate of 0.001 for 20 epochs. Note that ``relu" activation is used for the hidden and output layers. L2 regularization with tuning parameter 0.0001 is also used to optimize the loss for the NCF model. 

We then study the use of the learned user embeddings to suggest academic concentrations by training a logistic regression classifier.  We train the multi-class logistic regression model by minimizing a multinomial loss that fits across the entire probability distribution using  stochastic average gradient descent in batch mode with a learning rate of 0.001 for 500 epochs. We further use L2 regularization with tuning parameter 0.0001 in the logistic regression model. 

To gender-debias the recommendation, we add a de-biasing step prior to applying logistic regression. Our debiasing approach adapts a  recent work on attenuating bias in word vectors~\cite{dev2019attenuating}. Since traditional word embeddings are usually trained on massive text data, they inherit some of the human racial and gender biases from the data, as demonstrated by this well-known example, in which vector arithmetic on the embeddings solves an analogy task~\cite{buono}:

\begin{center}
{\tt doctor} $-$ {\tt man} + {\tt woman} = {\tt nurse} .
\end{center}

The user embeddings we have trained experience a very similar issue.  Let $p_u$ denote the embedding of a user, and let $v_B$, which is a unit vector in the same embedding space, denote the global gender bias in our system. We then debias $p_u$ by removing $p_u$'s projection on the bias vector $v_B$:

\begin{equation}
    p_{u}'= p_{u} - (p_{u}\cdot v_{B})v_{B} \mbox{ .}
\end{equation}

The question is how to find $v_B$. We consider $v_{female}$, given below, is the representation of an average female user:
$$v_{female} = \frac{1}{n_f}(f_1 + f_2 + \cdots + f_{n_f})$$ where $f_1,f_2,\cdots,f_{|n_f|}$ are the embeddings of female users. We define $v_{male}$ in the same way. This allows us to derive the overall gender bias vector as:
\[v_B=\frac{v_{female}-v_{male}}{\|v_{female}-v_{male}\|} \mbox{ .}\]
%
%
In the recommendation phase, the objective is to suggest top-$N$ academic concentrations to a new user based on the user's preference/interests indicated on Facebook. First, we construct the user embedding of the new user by analyzing the liked items of the user using the pre-trained NCF model. Then the embedding of the new user is used as input features in the pre-trained logistic regression classifier directly. For the gender-debiased system, we dropped the intercept terms during prediction to further remove popularity bias~\cite{steck2011item}. Finally, the logistic regression model recommends top-$N$ academic concentrations computed by sorting the probabilities of the 1380 concentrations for a given a user.

\section{Offline Evaluation}
\begin{table*}[t]
	\centering
	\small
    \begin{tabular}{lccccccc}
    \toprule
     
  & NDCG@$3\uparrow$ & NDCG@$10\uparrow$  & NDCG@$20\uparrow$ & $U_{PAR}\downarrow$ \\ \midrule
    GenderAware                      & 0.0009  & 0.005    & 0.007    & 1.1445     \\
     
    GenderDebiased   & \textbf{0.0050}    & \textbf{0.010}  & \textbf{0.013}   & \textbf{1.1188}    \\
    \bottomrule
    \end{tabular}
	\caption{Offline Evaluation on the \emph{Facebook} dataset. Higher Values are better for NDCG; lower values are better for  $U_{par}$. } 
	\label{tab:offlineevaluation}
\end{table*}
To compare the performance of the gender-debiased recommender with a gender-aware recommender, we implemented two variations of the same system. The \textbf{gender-aware system} makes career recommendations based on the choices by the people of the same gender (e.g., recommending to girls based on the career choices of other girls) while the \textbf{gender-debiased system} employs the linear projection-based gender de-biasing strategy to systematically remove gender stereotypes from user embeddings. 
\subsection{Evaluation Metrics}
We employ the following commonly used performance measures to evaluate the accuracy and fairness of each system.

\paragraph*{Normalized discounted cumulative
gain at K (NDCG@K)}
This is a well-known metric for assessing the quality of a {\it ranked} list of results (recommendations)~\cite{he2015trirank}. Since in our case there is only one ground truth recommendation (each Facebook user can only declare one career concentration), the metric can be simplified as:
\[NDCG@K=\sum_{i=1}^{N} \frac{relevance_K(n_i)}{\log_2(i+1)}\]
where $n_i$ is the $i$-th recommendation and
\begin{equation}
  relevance_K(n_i)=
  \begin{cases}
    1 & \text{if $i\le K$ and $n_i$ is the ground truth,}\\
    0 & \text{overotherwise}.
  \end{cases}
\end{equation}

In general, the higher the NDCG score is, the higher the prediction accuracy is. 
%

\paragraph*{Non-parity unfairness ($U_{PAR}$)}
This metric is designed to evaluate the fairness of recommender systems~\cite{Yao2017FairCF}. It computes the absolute difference of the average ratings between two groups of users:
$$U_{PAR} = |E_g[y] - E_{\neg g}[y]|$$
where $E_g[y]$ is the average predicted score from one group of users, and $E_{\neg g}[y]$ is the average predicted score for the other group of users. In our case, we consider scores for $N$ career concentrations for male and female subjects. 
$$U_{PAR}=\frac{1}{N}\sum_{n=1}^{N}|E_{female}[y_n]-E_{male}[y_n]|$$
In general, the lower $U_{PAR}$ is, the fairer the system is. 

\subsection{Experimental Settings and Results}
To train an AI model to predict career concentrations, we need negative examples as well, that is, an academic concentration that is not a good fit for a user. We generate random pairs $(u, c)$ as negative training instances, where $c$ is any academic concentration not explicitly declared by $u$. Furthermore, we split the data and use 70\% of it for training and the remaining 30\% for evaluation. 

Table~\ref{tab:offlineevaluation} shows the evaluation results. Since the NDCG scores at position 3, 10 and 20 for the gender-debiased system are consistently higher than those for the gender-aware system, the gender-debiased system is considered more accurate than the gender-aware system. In addition, since the gender-debiased system also has lower $U_{PAR}$ score, it is considered fairer than the gender-aware system.

During the offline evaluation with well-established evaluation metrics for recommendation accuracy and fairness,  we have demonstrated that the gender-debiased recommender is fairer without any loss of prediction accuracy, a criteria frequently  used in the fair machine learning community to indicate the success of a fair AI algorithm.  In the following, we describe a user study to investigate  whether the gender-debiased career recommender can achieve the  desired outcome with intended users, especially when gender-bias still exists in our society and even the most fairness-conscious individuals may still hold unconscious bias. 


\section{Online User Study}
\begin{figure*}[t]
		\centering
		\includegraphics[width=.85\textwidth]{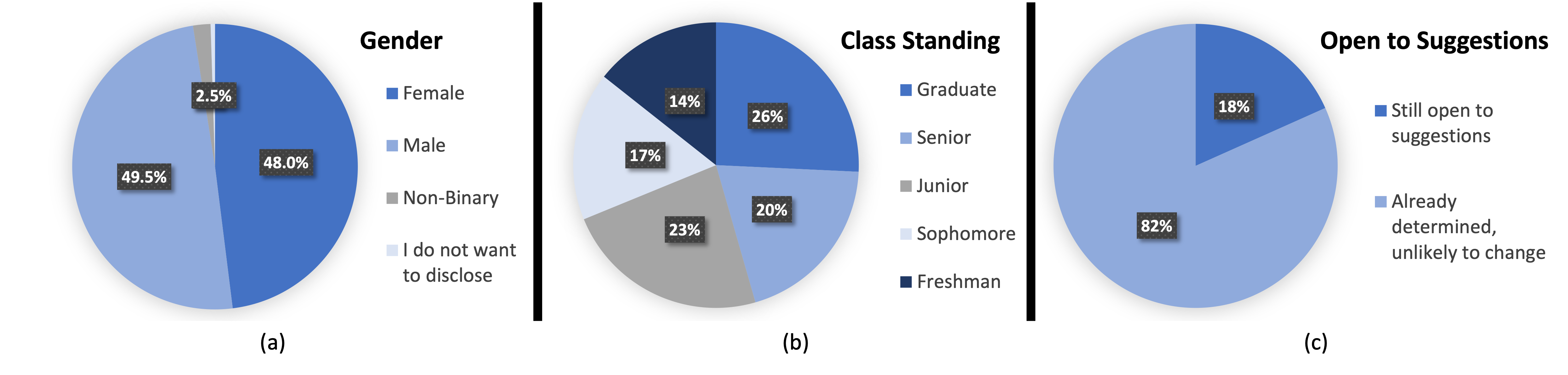}
		\caption{\small Participant Demographics: \emph{(a)} Gender, \emph{(b)} Academic Standings and \emph{(c)} Open to Suggestion.}
		\label{fig:Demographics}
\end{figure*} 


The goal of the user study was to investigate \emph{(a)} whether users prefer a gender-debiased career recommender over a gender-aware recommender, and \emph{(b)} whether their own belief/bias play a role in their preferences.

We adopted a between-subject design. We randomly assigned participants to use either a gender-debiased or a gender-aware career recommender (except for those who declare themselves gender non-binary or decline to disclose their gender, in which case the gender-debiased recommender was used). For participants assigned to the gender-aware recommender, we further assigned them to interact with either a "female" model or a "male" model, based on whether they were identified with the female or male gender.  Here, the "female (or male)" model means  a gender-aware career recommender trained on female (or male) data only.

We invited students from all majors at a mid-size university in the U.S. to participate in the online user study. Prior to the study, the survey protocol received the IRB approval, and participants  confirmed that they were 18 years or older and agreed to an informed consent before they could proceed. After taking the survey, each participant was entered in a raffle for \$50 Amazon Gift Cards. Overall, we received responses from 202 participants.  The entire survey took 5-10 minutes to complete. 

In the following, we describe the questionnaire used in the study. The questionnaire included five sections: demographics, user interests, beliefs about gender-bias in career choice, recommendation acceptance and general system usability.  
\subsection{Demographics}
We collected minimal personal information (such as gender) needed for the study.
\begin{enumerate}
    \item What gender do you identify as (female, male, non-binary, or do not want to disclose)?
    \item What is your class standing (freshman, sophomore, junior, senior, or graduate students)?
    \item How “set” is your choice of major / minor / concentration (still open to suggestions, or already determined, unlikely to Change )?
\end{enumerate}
As shown in Figure~\ref{fig:Demographics}, 48\% of the participants were females and 49.5\% males. Unlike the Facebook dataset used to train the backend algorithm, our online user study includes 2\%  participants who were gender non-binary and 0.5\% who did not want to disclose their genders. In terms of academic standing, 14.4\% were freshmen, 16.8\% were sophomore, 23.3\% were juniors, 19.8\% were seniors and 25.7\% were graduate students. Of all the participants, only 18.3\% were open to career suggestions. The rest were set on their chosen majors.

\subsection{User Interests and Preferences}
Users have varied interests and preferences. Accurately capturing their interests and preferences is critical to build a good recommender. In our work, we used a Facebook dataset to model user interests and preferences. The dataset stored $16K$ Facebook users' {\it likes} of $140K$ items. Given the large variety of the items, the set of items {\it liked} by a user could be a good representation of his or her interests and preferences.

Unfortunately, our participants were not among the $16K$ users in the Facebook dataset. In fact, many of them did  not even use Facebook. It was certainly impossible to ask our participants to indicate whether they like each of the $140K$ items in our dataset. In order to model our participants' interests and preferences, we first used a dimension reduction technique to group the 140K items  into a small number of categories. We then selected representative items from each category, and finally we asked our participants how they liked those representative items. 



Specifically, the Facebook dataset we used can be considered as a (sparse) user-item  matrix with $16,619$ rows (users) and $143,303$ columns (items). An entry is 1 if a person liked the item and 0 otherwise. We considered each item as a "word," and for each person, all the items he or she liked form a "document." We then performed a 100-topic Latent Dirichlet Allocation (LDA)~\cite{blei2003latent} analysis to automatically identify 100 latent topics in all the documents. Each of the latent topic was represented by a bag of "words", or in our case, a set of items (e.g., a latent topic related to high fantasy novels may contain representative items such as {\it The Lord of the Rings}, {\it The Hobbit}, {\it J.R.R. Tolkien}, {\it The Well at the World's End}, and {\it The Chronicles of Prydain}).    

From each of the 100 topics, we asked three volunteers to individually select one representative item from the top 10 items identified by LDA. We then picked the common items selected by the three volunteers to ensure most participants of our study are familiar with the items.  We finally decided on 48 well-known items, which are not too many to elicit a participant's interests in them during the user study. 

\subsection{Personal Belief on Gender Roles in Career Choice}
We asked two questions to help us understand a participant's beliefs about gender roles in career selection. 

\begin{enumerate}\setcounter{enumi}{3}
    \item \textbf{(Q-Stereotype)} Please indicate whether you agree with the following statement or not:  “A gender stereotype in career selection is undesirable since it limits women's and men's capacity to develop their personal abilities.”
    \item \textbf{(Q-DisparityPersonal)} Please indicate whether you agree with the following statement: “If I am a female, I do not want to choose a career that is male-dominated” {\it (for female participants)}, or “If I am a male, I do not want to choose a career that is female-dominated” {\it (for male participants)}.
\end{enumerate}
Both questions were rated on a 5-point Likert scale from strongly disagree to strongly agree.

\subsection{Recommendation Acceptance}
For each of the top-3 career recommendations, we asked a participant \emph{(a)} \textbf{(Q-Acceptance)} whether he/she will consider it as a possible future career choice (yes, no, I don't know)  and \emph{(b)} \textbf{(Q-DisparityCareer)} whether they perceive it to be a female- or male-dominated career or I do not know . The main difference between \textbf{(Q-DisparityCareer)} and \textbf{(Q-DisparityPersonal)} is that \textbf{(Q-DisparityPersonal)} expresses a general personal belief, while \textbf{(Q-DisparityCareer)} is a perception specific to a    major. 

\subsection{General System Usability}
We also asked about a participants' agreement with two general  usability-related statements: \emph{(a)} \textbf{(Q-UseAgain)} ``I would like to use a career recommendation system like this in the future,'' and \emph{(b)} \textbf{(Q-RecommendToOthers)} ``I would like to recommend the system to my friends if it is available.'' Both are rated on a 5-point Likert scale.

\section{Result Analysis}
We summarize the main findings of the user study, with a focus on \emph{(a)} user acceptance of gender-debiased versus gender-aware  recommendations, and \emph{(b)} whether a user's own belief or bias plays a role in the acceptance of a recommended career. 

\subsection{Summary of Recommendations by Each System}
\begin{figure*}[tb]
		\centerline{\includegraphics[width=.7\textwidth]{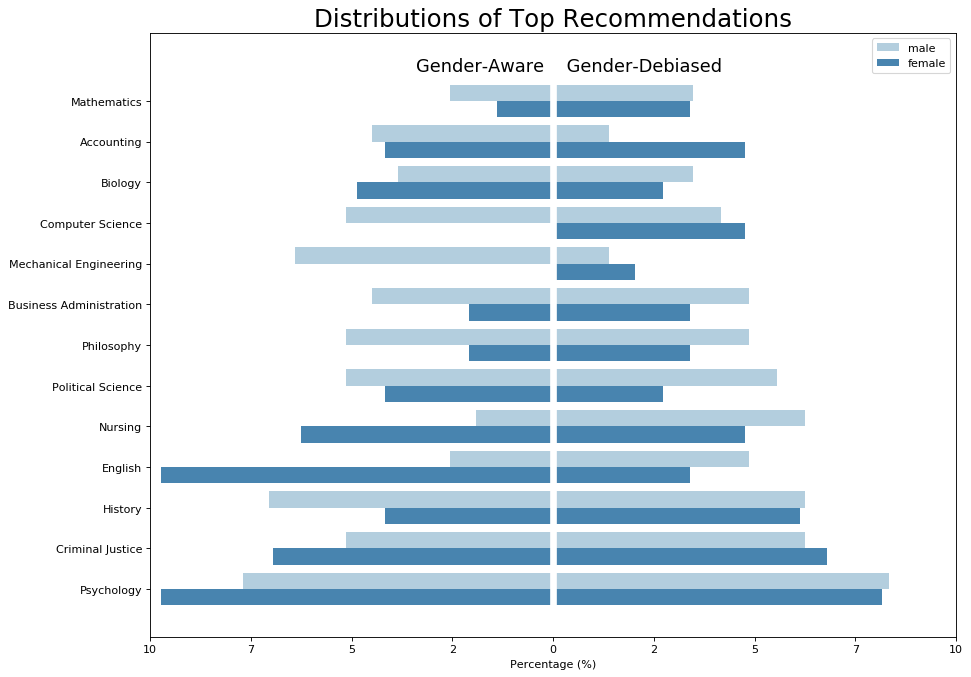}}
		\caption{\small Distributions of Career Recommendations by Gender from the \emph{Gender-Aware} and \emph{Gender-Debiased} Systems.}
		\label{fig:RecommendDisCombined}
\end{figure*}

The AI recommender gave each participant a recommendation of three concentrations (606 recommendations in total for 202 participants).
Figure ~\ref{fig:RecommendDisCombined} summarizes the recommendations by the gender-aware and the gender-debiased systems. It shows the top 13 most recommended career concentrations by both systems and the probabilities (on the x-axis) they are recommended 
for male and female participants respectively. As the left side of the chart shows, among all the career concentrations recommended by the gender-aware system, Psychology had 10\% chance of being recommended to females and 7.7\% chance of being recommended to males. Overall, the most frequently recommended careers by the gender-aware system for males were Psychology, Mechanical Engineering, History, Computer Science, and Criminal Justice; while the most frequently recommended careers for females were Psychology, English, Nursing, Biology and Accounting. Moreover, Computer Science and Mechanical Engineering were exclusively recommended to males by the gender-aware system. In contrast, the recommendations made by the gender-debiased system showed less gender stereotypes. As shown in the right chart of Figure~\ref{fig:RecommendDisCombined}, Computer Science was recommended to both males and females with similar probability (4.8\% versus 4.1\%). Based on this analysis, it seems the gender-debiased system is capable of mitigating  some existing gender biases in its recommendation. 

\subsection{Do People Prefer a Debiased Recommender?}
To test this, for each of the top 3 recommended concentrations, if a user indicated that they would  consider it as a possible future career choice, the system received 1 point. The system received 0 points if the user said “no” and 0.5 if the user said “I don’t know.” 
Based on an independent sample t-test, the mean acceptance score for the gender-debiased system was 0.279 while that for the gender-aware system was 0.372. The difference is statistically significant with $p<0.01$. Despite the results from the offline evaluation that showed the gender-debiased recommender was more fair while maintaining the same level of recommendation accuracy, users in general did not seem to prefer the recommendations made by the gender-debiased system more than those by the gender-aware system. In fact, the acceptance score for the gender-aware system was significantly higher than that for the gender-debiased system.

In the following, we try to explore whether a participant's own belief/bias plays a role in explaining the finding.

\subsection{Self-reported Belief and Acceptance}
\begin{figure}[t]
		\centerline{\includegraphics[width=.4\textwidth]{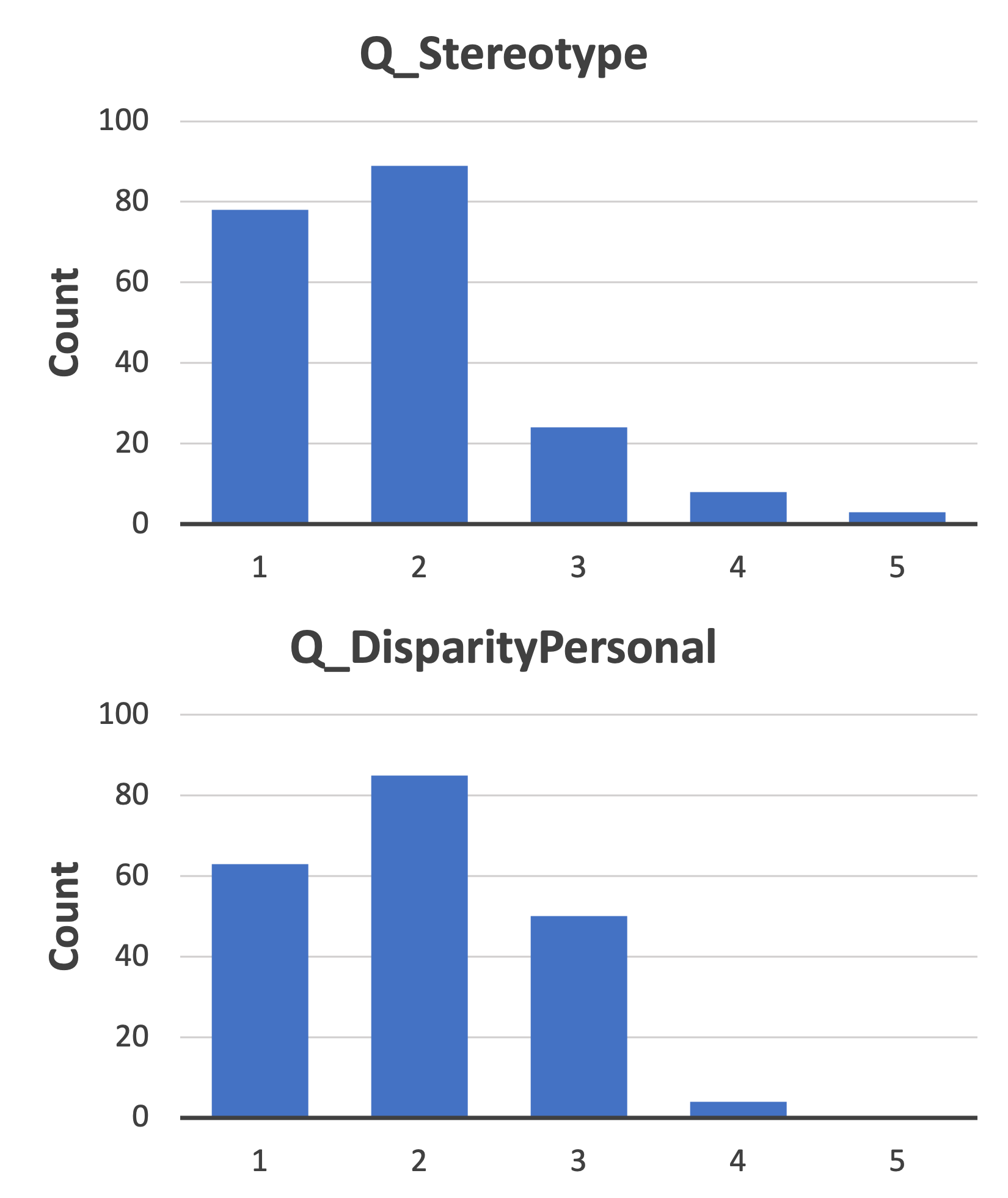}}
		\caption{\small Self-reported belief about the Gender Role in Career Choices. Q-Stereotypes (1: strongly agree or least biased and 5 strongly disagree or most biased) and Q-DisparityPersonal (1: Strongly disagree or least biased and 5 Strongly Agree or most biased)  
		}
		\label{fig:atititude}
\end{figure}

Here we focus on a participant's responses to Q-Stereotype and Q-DisparityPersonal. Both  responses were rated on a 5 point Likert scale, 5 being the most biased (strongly disagree with  Q-Stereotype or strongly agree with  Q-DisparityPersonal), 1 being the least (strongly agree with Q-Stereotype and strongly disagree with Q-DisparityPersonal) and 3 being neutral. Figure~\ref{fig:atititude} shows the distribution of the responses. The majority of the participants received a score of either "1" or "2",  Very few people scored more than 3. In fact, only 4\% of the participants scored 4 and 1.5\% scored 5 in Q-Stereotype. Only 2\% scored 4 and 0\% scored 5 in Q-DisparityPersonal. In summary, based on self-reported user responses to Q-Stereotype and Q-DisparityPersonal, only a  small number of the participants exhibited some degree of gender bias in college major selection. 

To test whether a participant's self-reported  belief impacts his/her acceptance of a recommendation, we employed a Generalized Linear Model (GLM) where the dependent variable was his/her acceptance score regarding a recommended major and the independent variables were his/her responses to Q-Stereotype or Q-DisparityPersonal. We also controlled the variation of demographics such as age, gender and academic standings as they could be confounders. 

The GLM results indicate that the main effect for Q-Stereotype on Q-Acceptance is not statistically significant($p<0.667$). In contrast, the main effect for Q-DisparityPersonal on Q-Acceptance is significant ($p<0.050$). 

Note the self-reported bias measures such as Q-Stereotype and Q-DisparityPerosnal may not accurately capture a person's true belief/bias. Prior research has demonstrated that social desirability bias is common in self-report surveys when the survey topics are sensitive (e.g., related to illegal acts such as drug use, income, ability and prejudice)~\cite{krumpal2013determinants}.  Due to social desirability concerns, there is a tendency for people to overreport socially desirable behaviors or attitudes and underreport  socially undesirable behaviors or attitudes.  Since gender bias is considered a sensitive topic, it is likely that our participants may have responded in a way that show less bias.

Mitigating social desirability in self-report surveys remains a challenging topic in social psychology as people differ in their tendency to engage in socially desirable responding~\cite{edwards1953relationship,fordyce1956social}. To overcome this problem, in the following, we measure a form of implicit bias based on perceived gender-disparity of an academic concentration. Since perceived gender-disparity of a major (e.g., to ask a person whether Computer Science is a male or female-dominated career) is a less personal and less sensitive topic, we expect to get a more accurate assessment of a participants' gender-related biases.


\subsection{Perceived Gender Conformity and Acceptance}

\begin{figure}[t]
		\centerline{\includegraphics[width=.3\textwidth]{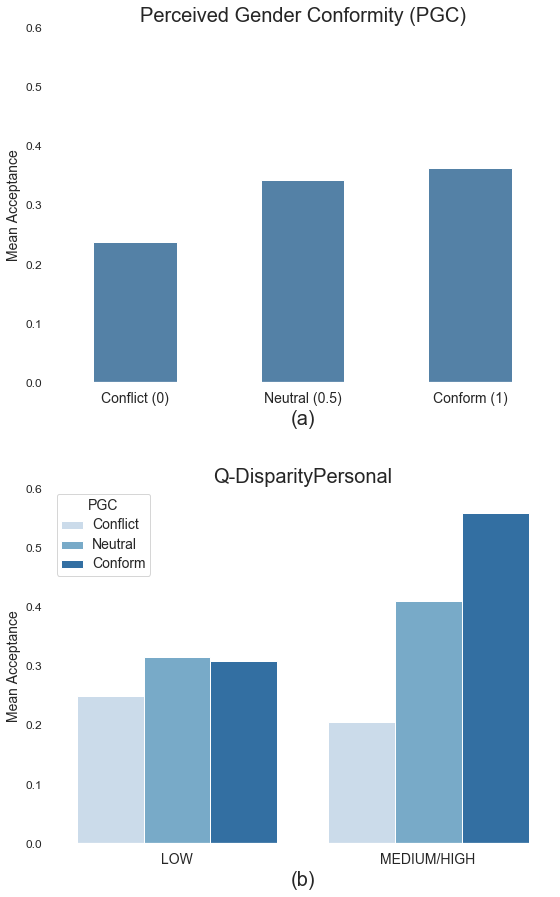}}
		\caption{\small \emph{(a)} The Relationship between User Acceptance (y-axis) and Perceived Gender Conformity (PGC) \emph{(b)} The interaction between Q-DisparityPersonal and PGC.
		}
		\label{fig:acceptancegenderconformity}
\end{figure}

We define perceived gender conformity (PGC), an implicit bias measure based on a participant's own gender and his/her responses to Q-DisparityCareer.  PGC is equal to "1" or "conform" if the perceived dominant gender of a major is consistent with the gender of the participant (e.g., a major is perceived to be male-dominated and the participant is a male or a major is perceived to be female-dominated and the participant is a female).  PGC will be "0" or "conflict" if the perceived dominant gender of the major  conflicts with the gender of the participant (e.g., the major is male-dominated and the participant is a female or the major is female-dominated and the participant is a male). For all the other cases (where participants answered "I don't know" to Q-DisparityCareer or the gender of the participant is non-binary or non-disclose), the value of PGC is assigned to "0.5" or "neutral". 

We built a GLM model to study the relation between a user's PGC and his/her acceptance of a recommended major. The dependent variable was Q-Acceptance and the independent variable was PGC. We controlled demographic factors such as age, gender and academic standing. Our analysis results show a positive correlation between PGC and user acceptance ($\beta=0.104$, p<0.050).  Figure~\ref{fig:acceptancegenderconformity}\emph{(a)} shows the average acceptance scores grouped by PGC. The mean acceptance score was the lowest (0.23) when the perceived "gender" of the recommended major differed  from the gender of the participant (PGC=0). In contrast, when they were the same (PGC=1), the mean acceptance score was the highest (0.364). When there was no perceived gender-disparity or the participant was gender non-binary or non-disclose (PGC=0.5), the mean acceptance score was in between(0.342). Since the acceptance gap between \textit{PGC=0} and \textit{PGC=0.5} was much larger than between \textit{PGC=0.5} and \textit{PGC=1}, the observed correlation seems mainly due to the avoidance of majors that were perceived to be dominated by the opposite gender. This may partially explain why our participants preferred the gender-debiased system less since it tried to overcome some of the gender stereotypes and was more likely to recommend majors dominated by the opposite sex. 

\subsection{Interaction Between Personal Belief and PGC}
We also studied whether there was any significant interaction effect between a user's personal belief  (Q-Stereotypes and Q-DisparityPersonal) and the perceived gender conformity of a major (PGC) on user acceptance. We performed two new GLM analyses where the dependent variable was Q-Acceptance and the independent variables were Q-Stereotype*PGC or Q-DisparityPersonal*PGC respectively. We also controlled for demographics such as age, gender and academic standing. Our results show that the interaction effect between a user's response to Q-Stereotype and PGC on Q-Acceptance was not significant ($p<0.9223$). But there was a marginally significant interaction effect between a user's responses to Q-DisparityPerosnal and PGC on user acceptance ($p<0.052$). 

To understand the interaction effect between Q-DisparityPersonal and PGC on user acceptance, we grouped the responses to Q-DisparityPersonal into LOW (those with a score of 1 or 2) and MEDIUM/HIGH (those with a score of 3, 4, and 5). Since there were very few people with a score of 4 or 5 (only 2\% of the participants), most of the people in the  Medium/High group had a score of 3. Figure~\ref{fig:acceptancegenderconformity}(b) shows the comparison between these two groups of people. When people did not mind selecting a major dominated by the opposite sex (in the Q-DisparityPersonal=LOW group), there wasn't much difference in their acceptance  of majors conflicting or conforming to their genders. In contrast, people in the MEDIUM/HIGH group showed high  preference to majors conforming to their genders (PGC=1) while avoiding majors conflicting with their genders (PGC=0).

\subsection{User Acceptance and Other Factors}
Among the demographics of a participant,  gender was found to be significantly correlated with recommendation acceptance ($P<0.017$); the correlation between age and recommendation acceptance was marginally significant ($p<0.063$). Academic standing was not significantly correlated with the acceptance ($p<0.911$). Moreover, among all the factors we tested, the one that was most predictive of whether someone would consider the recommended career a possible future career choice is whether someone was still open to career changes ($p<0.000$). When asked the question "How 'set' are you on your choice of major / minor / concentration?", $82\%$ chose "Already determined, unlikely to change", only $18\%$ chose "Still open to suggestions" (see Figure~\ref{fig:Demographics}\emph{c)}). Unsurprisingly, most people would not accept the recommended careers, regardless of the quality of the recommendations.  

\subsection{General System Usability}

\begin{figure}[t]
		\centerline{\includegraphics[width=.4\textwidth]{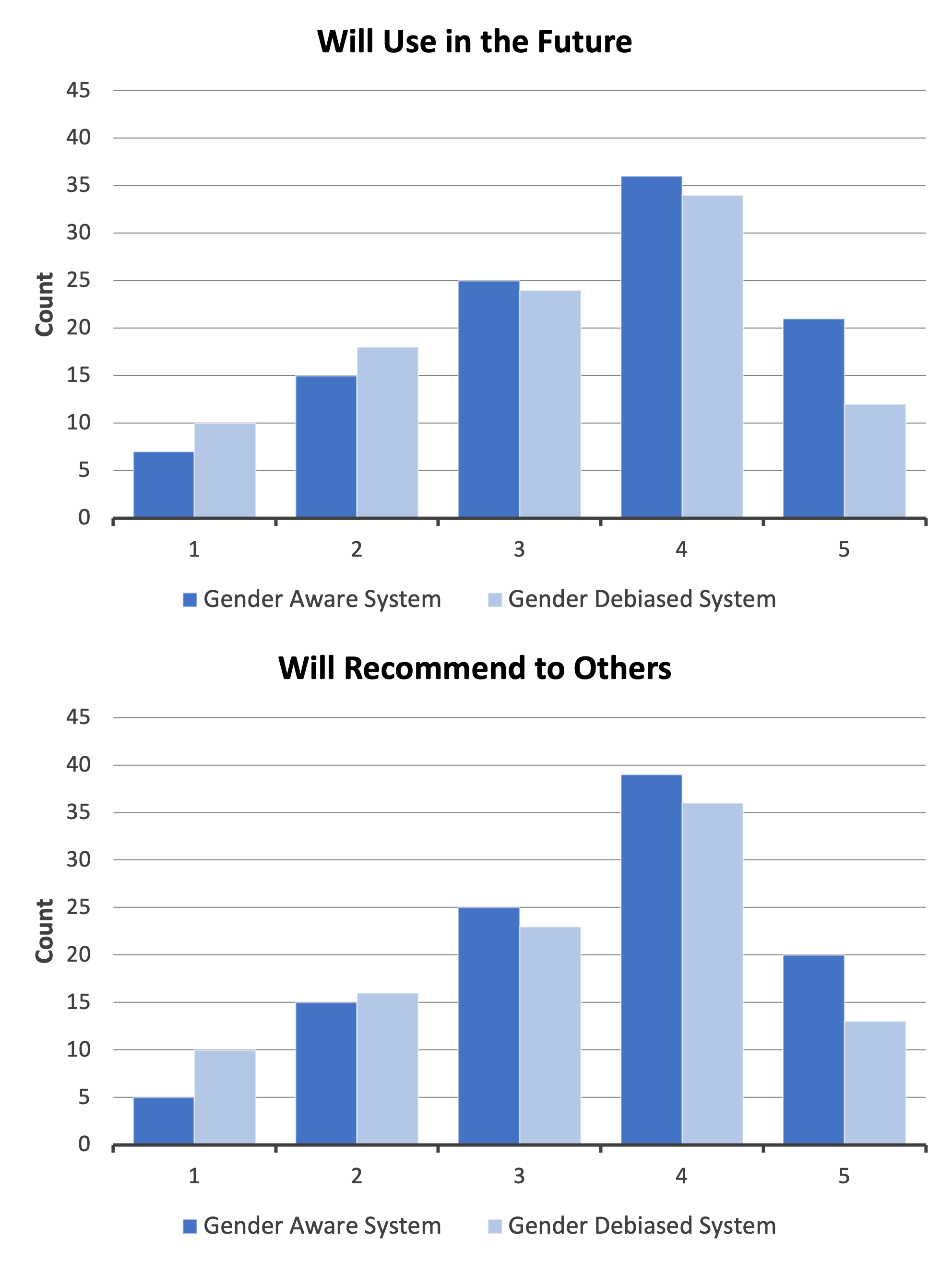}}
		\caption{\small Distributions of \emph{(a)} \emph{Q-UseAgain (left)} and \emph{(b)} \emph{Q-RecommendToOthers (right)}.
		}
		\label{fig:Usability}
\end{figure}
Finally, based on user responses to two general system usability questions (Q-UseAgain and Q-RecommendToOthers), our participants were generally positive about the systems. The mean score for Q-UseAgain was 3.34 and the mean for Q-RecommendToOthers was 3.40, both are better than neutral (3). Figure~\ref{fig:Usability} shows the response distributions by different systems (5 being the best). Since the response distributions for Q-UseAgain and Q-RecommendToOthers are very similar, they show a consistency between these two usability measures. In addition, since the darker  bars skewed more toward right, the mean usability scores for the gender-aware system are generally higher than those of the gender-debiased system (for Q-UseAgain, the mean is 3.20 for the gender-debiased system and 3.47 for the gender-aware system; for Q-RecommendToOthers the mean is 3.27 for the gender-debiased system and 3.52 for the gender-aware system). The differences are marginally significant (for Q-UseAgain, $p<0.095$, for Q-RecommendToOthers: $p<0.093$). This result is consistent with the Q-Acceptance-based evaluation measure, which confirms our prior finding: the study participants preferred the gender-aware system more than the gender de-biased system.  

\section{Discussion}
In this section, we discuss the findings we have discovered in this work, its implications on fair AI system design, and the limitations of our current study which could be addressed in future work. 

\subsection{What Have We Discovered?}
Much effort in the AI and HCI community has focused on identifying and removing AI bias using algorithmic or design-based solutions. What we have discovered is that our \emph{participants prefer biased recommendations over debiased ones}, despite both methods achieving similar prediction accuracy.  \emph{For debiasing algorithms such as our fair career recommender to move the needle on equity in the real world, we may also need to find ways to ``nudge'' users to accept debiased recommendations.}  

To understand why participants did not prefer fair recommendations, our study results suggest that their unconscious bias may play a role. Our participants seemed to avoid careers that are dominated by the opposite sex. Similar results were found in  previous research~\cite{stroeher1994sixteen} where, even as kindergartners, girls would select mostly traditionally female careers  and avoid traditionally male careers. In other words, the bias is so deeply ingrained that people subconsciously shun a career dominated by the opposite gender regardless whether it matches their interests and skills.  Societal bias may also play a role in the participants' preferences for gendered recommendation.  
Even supposing that a person is unbiased, they may still make a career choice that conforms to ``social norms'' if they believe that they might otherwise be disadvantaged in their career growth or subjected to discrimination on the job. 

Since systemic bias and prejudice due to humans are root causes of inequities in our society \cite{crenshaw1989demarginalizing} and hence in data, without addressing the human side of the issue, fair AI systems may not produce their intended societal impacts. Beyond addressing algorithmic issues, the AI and the HCI communities need to devote more attention to \emph{developing technologies that can help humans identify and overcome their own biases}.

\subsection{Implications on Fair AI Research}


We believe in addition to debiasing algorithms, bias/fairness explanation and persuasion technology may play important roles in helping people overcome their biases.  
Educating people by explaining fairness/bias as well as its societal impact may help raise awareness and slowly nudge people toward a fairer society.

\subsection{Limitations of the current study}
One limitation of the current study is that over 80\% of the participants are not open to new career suggestions, which significantly limits the statistical power of our analysis. As most of the participants are university students and have decided a career concentration, it would be better if most participants have not decided a career, and thus could benefit more from the recommendations. A future possibility is to revise the current IRB protocol to recruit minors (e.g., high school students) for the study.

\section{Conclusions}
In this paper, we demonstrated that it is not sufficient to simply rely on AI/ML debiasing algorithms to achieve the desired societal outcomes that the field of fair ML aims to produce, at least in the context of gender-debiased career recommendations. In fact, we found that on average our participants did not prefer career recommendations from a gender-debiased system, even though the system was just as accurate as the corresponding gender-aware system and participants generally self-reported that gender stereotypes in career selections are undesirable. Our results suggest that participants' own beliefs and unconscious biases are contributing factors to their acceptance of the AI recommendations (e.g., participants tended to avoid careers dominated by the opposite sex). 
To improve real-world equity in careers via a debiased AI fairness system, it may be necessary to counter human bias as well as AI bias.  

Going forward, it would be valuable to repeat our experiments in other problem domains, and with other protected dimensions such as race and disability status.  To ensure that the impacts of fair AI technologies fulfil their potential benefits to society, more research on human-fair AI interaction, an understudied area, is urgently needed.

\begin{quote}
\begin{small}
\bibliography{acmart}

\begin{thebibliography}{72}
\providecommand{\natexlab}[1]{#1}
\providecommand{\url}[1]{\texttt{#1}}
\providecommand{\urlprefix}{URL }
\expandafter\ifx\csname urlstyle\endcsname\relax
  \providecommand{\doi}[1]{doi:\discretionary{}{}{}#1}\else
  \providecommand{\doi}{doi:\discretionary{}{}{}\begingroup
  \urlstyle{rm}\Url}\fi

\bibitem[{Agarwal et~al.(2018)Agarwal, Beygelzimer, Dud{\'\i}k, Langford, and
  Wallach}]{agarwal2018reductions}
Agarwal, A.; Beygelzimer, A.; Dud{\'\i}k, M.; Langford, J.; and Wallach, H.
  2018.
\newblock A reductions approach to fair classification.
\newblock \emph{arXiv preprint arXiv:1803.02453} .

\bibitem[{Ali et~al.(2019)Ali, Sapiezynski, Bogen, Korolova, Mislove, and
  Rieke}]{ali2019discrimination}
Ali, M.; Sapiezynski, P.; Bogen, M.; Korolova, A.; Mislove, A.; and Rieke, A.
  2019.
\newblock Discrimination through Optimization: How Facebook's Ad Delivery Can
  Lead to Biased Outcomes.
\newblock \emph{Proceedings of the ACM on Human-Computer Interaction} 3(CSCW):
  1--30.

\bibitem[{Angwin et~al.(2016)Angwin, Larson, Mattu, and
  Kirchner}]{angwin2016machine}
Angwin, J.; Larson, J.; Mattu, S.; and Kirchner, L. 2016.
\newblock Machine bias: There’s software used across the country to predict
  future criminals. and it’s biased against blacks.
\newblock \emph{ProPublica, May} 23.

\bibitem[{authors(2019)}]{Anonymous1}
authors, A. 2019.
\newblock Details omitted for double-blind reviewing.
\newblock Anonymous Non-archival Workshop (extended abstract). Please see the
  paper as an anonymized document in the supplement material.

\bibitem[{Bandura(1977)}]{bandura1977self}
Bandura, A. 1977.
\newblock Self-efficacy: toward a unifying theory of behavioral change.
\newblock \emph{Psychological review} 84(2): 191.

\bibitem[{Bandura(1978)}]{bandura1978reflections}
Bandura, A. 1978.
\newblock Reflections on self-efficacy.
\newblock \emph{Advances in behaviour research and therapy} 1(4): 237--269.

\bibitem[{Barbosa and Chen(2019)}]{barbosa2019rehumanized}
Barbosa, N.~M.; and Chen, M. 2019.
\newblock Rehumanized crowdsourcing: a labeling framework addressing bias and
  ethics in machine learning.
\newblock In \emph{Proceedings of the 2019 CHI Conference on Human Factors in
  Computing Systems}, 1--12.

\bibitem[{Barocas and Selbst(2016)}]{barocas2016big}
Barocas, S.; and Selbst, A. 2016.
\newblock Big data's disparate impact.
\newblock \emph{Cal. L. Rev.} 104: 671.

\bibitem[{Bellamy et~al.(2019)Bellamy, Dey, Hind, Hoffman, Houde, Kannan,
  Lohia, Martino, Mehta, Mojsilovic, Nagar, Ramamurthy, Richards, Saha,
  Sattigeri, Singh, Varshney, and Zhang}]{Bellamy19}
Bellamy, R. K.~E.; Dey, K.; Hind, M.; Hoffman, S.~C.; Houde, S.; Kannan, K.;
  Lohia, P.; Martino, J.; Mehta, S.; Mojsilovic, A.; Nagar, S.; Ramamurthy,
  K.~N.; Richards, J.~T.; Saha, D.; Sattigeri, P.; Singh, M.; Varshney, K.~R.;
  and Zhang, Y. 2019.
\newblock AI Fairness 360: An extensible toolkit for detecting and mitigating
  algorithmic bias.
\newblock \emph{IBM J. Res. Dev.} 63(4/5): 4:1--4:15.

\bibitem[{Blei, Ng, and Jordan(2003)}]{blei2003latent}
Blei, D.~M.; Ng, A.~Y.; and Jordan, M.~I. 2003.
\newblock Latent dirichlet allocation.
\newblock \emph{Journal of machine Learning research} 3(Jan): 993--1022.

\bibitem[{Bolukbasi et~al.(2016)Bolukbasi, Chang, Zou, Saligrama, and
  Kalai}]{bolukbasi2016man}
Bolukbasi, T.; Chang, K.-W.; Zou, J.; Saligrama, V.; and Kalai, A. 2016.
\newblock Man is to Computer Programmer as Woman is to Homemaker? {D}ebiasing
  Word Embeddings.
\newblock In \emph{Advances in NeurIPS}.

\bibitem[{Buonocore(2019)}]{buono}
Buonocore, T. 2019.
\newblock Man is to doctor as woman is to nurse: the gender bias of word
  embeddings.
\newblock
  \url{https://towardsdatascience.com/gender-bias-word-embeddings-76d9806a0e17}.

\bibitem[{Campolo et~al.(2017)Campolo, Sanfilippo, Whittaker, K.~Crawford, and
  Barocas}]{campolo2017ai}
Campolo, A.; Sanfilippo, M.; Whittaker, M.; K.~Crawford, A.~S.; and Barocas, S.
  2017.
\newblock \emph{{AI} {N}ow 2017 Symposium Report}.
\newblock AI Now.

\bibitem[{Chen et~al.(2020)Chen, Xu, Liu, Guo, Liu, Tong, Akkiraju, and
  Carroll}]{chen2020general}
Chen, J.; Xu, A.; Liu, Z.; Guo, Y.; Liu, X.; Tong, Y.; Akkiraju, R.; and
  Carroll, J.~M. 2020.
\newblock A General Methodology to Quantify Biases in Natural Language Data.
\newblock In \emph{Extended Abstracts of the 2020 CHI Conference on Human
  Factors in Computing Systems}, 1--9.

\bibitem[{Correll(2001)}]{correll2001gender}
Correll, S.~J. 2001.
\newblock Gender and the career choice process: The role of biased
  self-assessments.
\newblock \emph{American journal of Sociology} 106(6): 1691--1730.

\bibitem[{Cramer et~al.(2019)Cramer, Garcia-Gathright, Reddy, Springer, and
  Takeo~Bouyer}]{cramer2019translation}
Cramer, H.; Garcia-Gathright, J.; Reddy, S.; Springer, A.; and Takeo~Bouyer, R.
  2019.
\newblock Translation, tracks \& data: an algorithmic bias effort in practice.
\newblock In \emph{Extended Abstracts of the 2019 CHI Conference on Human
  Factors in Computing Systems}, 1--8.

\bibitem[{Crenshaw(1989)}]{crenshaw1989demarginalizing}
Crenshaw, K. 1989.
\newblock Demarginalizing the intersection of race and sex: A black feminist
  critique of antidiscrimination doctrine, feminist theory and antiracist
  politics.
\newblock \emph{U. Chi. Legal F.} 139--167.

\bibitem[{Cryan et~al.(2020)Cryan, Tang, Zhang, Metzger, Zheng, and
  Zhao}]{cryan2020detecting}
Cryan, J.; Tang, S.; Zhang, X.; Metzger, M.; Zheng, H.; and Zhao, B.~Y. 2020.
\newblock Detecting Gender Stereotypes: Lexicon vs. Supervised Learning
  Methods.
\newblock In \emph{Proceedings of the 2020 CHI Conference on Human Factors in
  Computing Systems}, 1--11.

\bibitem[{Dastin(2018)}]{dastin2018amazon}
Dastin, J. 2018.
\newblock Amazon scraps secret {AI} recruiting tool that showed bias against
  women.
\newblock \emph{Reuters} \urlprefix\url{https://www.reuters.com/article/
  us-amazon-com-jobs-automation-insight/
  amazon-scraps-secret-ai-recruiting-tool-that-showed-bias-against-women-idUSKCN1MK08G}.

\bibitem[{Dev and Phillips(2019)}]{dev2019attenuating}
Dev, S.; and Phillips, J. 2019.
\newblock Attenuating Bias in Word Vectors.

\bibitem[{D{\'\i}az et~al.(2018)D{\'\i}az, Johnson, Lazar, Piper, and
  Gergle}]{diaz2018addressing}
D{\'\i}az, M.; Johnson, I.; Lazar, A.; Piper, A.~M.; and Gergle, D. 2018.
\newblock Addressing age-related bias in sentiment analysis.
\newblock In \emph{Proceedings of the 2018 CHI Conference on Human Factors in
  Computing Systems}, 1--14.

\bibitem[{Dodge et~al.(2019)Dodge, Liao, Zhang, Bellamy, and Dugan}]{Dodge2019}
Dodge, J.; Liao, Q.~V.; Zhang, Y.; Bellamy, R. K.~E.; and Dugan, C. 2019.
\newblock Explaining Models: An Empirical Study of How Explanations Impact
  Fairness Judgment.
\newblock IUI '19, 275–285. New York, NY, USA: Association for Computing
  Machinery.
\newblock ISBN 9781450362726.

\bibitem[{Dwork et~al.(2012)Dwork, Hardt, Pitassi, Reingold, and
  Zemel}]{dwork2012fairness}
Dwork, C.; Hardt, M.; Pitassi, T.; Reingold, O.; and Zemel, R. 2012.
\newblock Fairness through awareness.
\newblock In \emph{Proceedings of ITCS}, 214--226. ACM.

\bibitem[{Edwards(1953)}]{edwards1953relationship}
Edwards, A.~L. 1953.
\newblock The relationship between the judged desirability of a trait and the
  probability that the trait will be endorsed.
\newblock \emph{Journal of Applied Psychology} 37(2): 90.

\bibitem[{Fordyce(1956)}]{fordyce1956social}
Fordyce, W.~E. 1956.
\newblock Social desirability in the MMPI.
\newblock \emph{Journal of Consulting Psychology} 20(3): 171.

\bibitem[{Foulds et~al.(2020{\natexlab{a}})Foulds, Islam, Keya, and
  Pan}]{foulds2020bayesian}
Foulds, J.~R.; Islam, R.; Keya, K.~N.; and Pan, S. 2020{\natexlab{a}}.
\newblock Bayesian Modeling of Intersectional Fairness: The Variance of Bias.
\newblock In \emph{Proceedings of the 2020 SIAM International Conference on
  Data Mining}, 424--432. SIAM.

\bibitem[{Foulds et~al.(2020{\natexlab{b}})Foulds, Islam, Keya, and
  Pan}]{foulds2020intersectional}
Foulds, J.~R.; Islam, R.; Keya, K.~N.; and Pan, S. 2020{\natexlab{b}}.
\newblock An intersectional definition of fairness.
\newblock In \emph{2020 IEEE 36th International Conference on Data Engineering
  (ICDE)}, 1918--1921. IEEE.

\bibitem[{Gajane and Pechenizkiy(2017)}]{gajane2017formalizing}
Gajane, P.; and Pechenizkiy, M. 2017.
\newblock On formalizing fairness in prediction with machine learning.
\newblock \emph{arXiv preprint arXiv:1710.03184} .

\bibitem[{Glick and Fiske(1999)}]{glick1999gender}
Glick, P.; and Fiske, S.~T. 1999.
\newblock Gender, power dynamics, and social interaction.
\newblock \emph{Revisioning gender} 5: 365--398.

\bibitem[{Hardt, Price, and Srebro(2016)}]{hardt2016equality}
Hardt, M.; Price, E.; and Srebro, N. 2016.
\newblock Equality of opportunity in supervised learning.
\newblock In \emph{Advances in NeurIPS}, 3315--3323.

\bibitem[{He et~al.(2015)He, Chen, Kan, and Chen}]{he2015trirank}
He, X.; Chen, T.; Kan, M.-Y.; and Chen, X. 2015.
\newblock Trirank: Review-aware explainable recommendation by modeling aspects.
\newblock In \emph{Proceedings of the 24th ACM International on Conference on
  Information and Knowledge Management}, 1661--1670.

\bibitem[{He et~al.(2017)He, Liao, Zhang, Nie, Hu, and Chua}]{he2017neural}
He, X.; Liao, L.; Zhang, H.; Nie, L.; Hu, X.; and Chua, T. 2017.
\newblock Neural Collaborative Filtering.
\newblock \emph{CoRR} abs/1708.05031.
\newblock \urlprefix\url{http://arxiv.org/abs/1708.05031}.

\bibitem[{Hill and Giles(2014)}]{hill2014career}
Hill, E.~J.; and Giles, J.~A. 2014.
\newblock Career decisions and gender: the illusion of choice?
\newblock \emph{Perspectives on medical education} 3(3): 151--154.

\bibitem[{Hou et~al.(2017)Hou, Lampe, Bulinski, and Prescott}]{hou2017factors}
Hou, Y.; Lampe, C.; Bulinski, M.; and Prescott, J.~J. 2017.
\newblock Factors in Fairness and Emotion in Online Case Resolution Systems.
\newblock In \emph{Proceedings of the 2017 CHI Conference on Human Factors in
  Computing Systems}, 2511--2522.

\bibitem[{Hyde, Fennema, and Lamon(1990)}]{hyde1990gender}
Hyde, J.~S.; Fennema, E.; and Lamon, S.~J. 1990.
\newblock Gender differences in mathematics performance: A meta-analysis.
\newblock \emph{Psychological bulletin} 107(2): 139.

\bibitem[{Johnson et~al.(2017)Johnson, McMahon, Sch{\"o}ning, and
  Hecht}]{johnson2017effect}
Johnson, I.; McMahon, C.; Sch{\"o}ning, J.; and Hecht, B. 2017.
\newblock The Effect of Population and" Structural" Biases on Social
  Media-based Algorithms: A Case Study in Geolocation Inference Across the
  Urban-Rural Spectrum.
\newblock In \emph{Proceedings of the 2017 CHI conference on Human Factors in
  Computing Systems}, 1167--1178.

\bibitem[{Kay, Matuszek, and Munson(2015)}]{kay2015unequal}
Kay, M.; Matuszek, C.; and Munson, S.~A. 2015.
\newblock Unequal representation and gender stereotypes in image search results
  for occupations.
\newblock In \emph{Proceedings of the 33rd Annual ACM Conference on Human
  Factors in Computing Systems}, 3819--3828.

\bibitem[{Kosinski et~al.(2015)Kosinski, Matz, Gosling, Popov, and
  Stillwell}]{kosinski2015facebook}
Kosinski, M.; Matz, S.~C.; Gosling, S.~D.; Popov, V.; and Stillwell, D. 2015.
\newblock Facebook as a research tool for the social sciences: Opportunities,
  challenges, ethical considerations, and practical guidelines.
\newblock \emph{American Psychologist} 70(6): 543.

\bibitem[{Krumpal(2013)}]{krumpal2013determinants}
Krumpal, I. 2013.
\newblock Determinants of social desirability bias in sensitive surveys: a
  literature review.
\newblock \emph{Quality \& Quantity} 47(4): 2025--2047.

\bibitem[{Kusner et~al.(2017)Kusner, Loftus, Russell, and Silva}]{Kusner2017}
Kusner, M.~J.; Loftus, J.; Russell, C.; and Silva, R. 2017.
\newblock Counterfactual Fairness.
\newblock In Guyon, I.; Luxburg, U.~V.; Bengio, S.; Wallach, H.; Fergus, R.;
  Vishwanathan, S.; and Garnett, R., eds., \emph{Advances in Neural Information
  Processing Systems 30}, 4066--4076.

\bibitem[{Leaper and Starr(2019)}]{leaper2019helping}
Leaper, C.; and Starr, C.~R. 2019.
\newblock Helping and hindering undergraduate women’s STEM motivation:
  experiences With STEM encouragement, STEM-related gender bias, and sexual
  harassment.
\newblock \emph{Psychology of Women Quarterly} 43(2): 165--183.

\bibitem[{Lent, Brown, and Hackett(1994)}]{lent1994toward}
Lent, R.~W.; Brown, S.~D.; and Hackett, G. 1994.
\newblock Toward a unifying social cognitive theory of career and academic
  interest, choice, and performance.
\newblock \emph{Journal of vocational behavior} 45(1): 79--122.

\bibitem[{Lockwood(2006)}]{lockwood2006someone}
Lockwood, P. 2006.
\newblock “Someone like me can be successful”: Do college students need
  same-gender role models?
\newblock \emph{Psychology of Women Quarterly} 30(1): 36--46.

\bibitem[{Madaio et~al.(2020)Madaio, Stark, Wortman~Vaughan, and
  Wallach}]{madaio2020co}
Madaio, M.~A.; Stark, L.; Wortman~Vaughan, J.; and Wallach, H. 2020.
\newblock Co-Designing Checklists to Understand Organizational Challenges and
  Opportunities around Fairness in AI.
\newblock In \emph{Proceedings of the 2020 CHI Conference on Human Factors in
  Computing Systems}, 1--14.

\bibitem[{Matz et~al.(2017)Matz, Kosinski, Nave, and
  Stillwell}]{matz2017psychological}
Matz, S.~C.; Kosinski, M.; Nave, G.; and Stillwell, D.~J. 2017.
\newblock Psychological targeting as an effective approach to digital mass
  persuasion.
\newblock \emph{Proceedings of the national academy of sciences} 114(48):
  12714--12719.

\bibitem[{McDonald and Pan(2020)}]{McDonald2020}
McDonald, N.; and Pan, S. 2020.
\newblock Intersectional AI: A Study of How Information Science Students Think
  about Ethics and Their Impact.
\newblock \emph{Proc. ACM Hum.-Comput. Interact.} 4(CSCW2).

\bibitem[{McQuaid and Bond(2004)}]{mcquaid2004gender}
McQuaid, R.; and Bond, S. 2004.
\newblock Gender stereotyping in career choice .

\bibitem[{Meece et~al.(1982)Meece, Parsons, Kaczala, and Goff}]{meece1982sex}
Meece, J.~L.; Parsons, J.~E.; Kaczala, C.~M.; and Goff, S.~B. 1982.
\newblock Sex differences in math achievement: Toward a model of academic
  choice.
\newblock \emph{Psychological Bulletin} 91(2): 324.

\bibitem[{Mehrabi et~al.(2019)Mehrabi, Morstatter, Saxena, Lerman, and
  Galstyan}]{mehrabi2019survey}
Mehrabi, N.; Morstatter, F.; Saxena, N.; Lerman, K.; and Galstyan, A. 2019.
\newblock A survey on bias and fairness in machine learning.
\newblock \emph{arXiv preprint arXiv:1908.09635} .

\bibitem[{Metaxa-Kakavouli et~al.(2018)Metaxa-Kakavouli, Wang, Landay, and
  Hancock}]{metaxa2018gender}
Metaxa-Kakavouli, D.; Wang, K.; Landay, J.~A.; and Hancock, J. 2018.
\newblock Gender-inclusive design: Sense of belonging and bias in web
  interfaces.
\newblock In \emph{Proceedings of the 2018 CHI Conference on Human Factors in
  Computing Systems}, 1--6.

\bibitem[{Miller and Hosanagar(2010)}]{Miller2019}
Miller, A.~P.; and Hosanagar, K. 2010.
\newblock How Targeted Ads and Dynamic Pricing Can Perpetuate Bias.
\newblock \emph{Harvard Business Review} .

\bibitem[{Noble(2018)}]{noble2018algorithms}
Noble, S. 2018.
\newblock \emph{Algorithms of Oppression: How Search Engines Reinforce Racism}.
\newblock NYU Press.

\bibitem[{Otterbacher, Bates, and Clough(2017)}]{otterbacher2017competent}
Otterbacher, J.; Bates, J.; and Clough, P. 2017.
\newblock Competent men and warm women: Gender stereotypes and backlash in
  image search results.
\newblock In \emph{Proceedings of the 2017 chi conference on human factors in
  computing systems}, 6620--6631.

\bibitem[{Salminen et~al.(2020)Salminen, Jung, Chowdhury, and
  Jansen}]{salminen2020analyzing}
Salminen, J.; Jung, S.-g.; Chowdhury, S.; and Jansen, B.~J. 2020.
\newblock Analyzing Demographic Bias in Artificially Generated Facial Pictures.
\newblock In \emph{Extended Abstracts of the 2020 CHI Conference on Human
  Factors in Computing Systems}, 1--8.

\bibitem[{Salminen, Jung, and Jansen(2019)}]{salminen2019detecting}
Salminen, J.; Jung, S.-G.; and Jansen, B.~J. 2019.
\newblock Detecting Demographic Bias in Automatically Generated Personas.
\newblock In \emph{Extended Abstracts of the 2019 CHI Conference on Human
  Factors in Computing Systems}, 1--6.

\bibitem[{Sarwar et~al.(2001)Sarwar, Karypis, Konstan, and
  Riedl}]{sarwar2001item}
Sarwar, B.; Karypis, G.; Konstan, J.; and Riedl, J. 2001.
\newblock Item-based collaborative filtering recommendation algorithms.
\newblock In \emph{Proceedings of the 10th international conference on World
  Wide Web}, 285--295.

\bibitem[{Schwartz et~al.(2013)Schwartz, Eichstaedt, Kern, Dziurzynski,
  Ramones, Agrawal, Shah, Kosinski, Stillwell, Seligman
  et~al.}]{schwartz2013personality}
Schwartz, H.~A.; Eichstaedt, J.~C.; Kern, M.~L.; Dziurzynski, L.; Ramones,
  S.~M.; Agrawal, M.; Shah, A.; Kosinski, M.; Stillwell, D.; Seligman, M.~E.;
  et~al. 2013.
\newblock Personality, gender, and age in the language of social media: The
  open-vocabulary approach.
\newblock \emph{PloS one} 8(9): e73791.

\bibitem[{Skinner, Brown, and Walsh(2020)}]{skinner2020children}
Skinner, Z.; Brown, S.; and Walsh, G. 2020.
\newblock Children of Color's Perceptions of Fairness in AI: An Exploration of
  Equitable and Inclusive Co-Design.
\newblock In \emph{Extended Abstracts of the 2020 CHI Conference on Human
  Factors in Computing Systems}, 1--8.

\bibitem[{Steck(2011)}]{steck2011item}
Steck, H. 2011.
\newblock Item popularity and recommendation accuracy.
\newblock In \emph{Proceedings of the fifth ACM conference on Recommender
  systems}, 125--132.

\bibitem[{Strengers et~al.(2020)Strengers, Qu, Xu, and
  Knibbe}]{strengers2020adhering}
Strengers, Y.; Qu, L.; Xu, Q.; and Knibbe, J. 2020.
\newblock Adhering, Steering, and Queering: Treatment of Gender in Natural
  Language Generation.
\newblock In \emph{Proceedings of the 2020 CHI Conference on Human Factors in
  Computing Systems}, 1--14.

\bibitem[{Stroeher(1994)}]{stroeher1994sixteen}
Stroeher, S.~K. 1994.
\newblock Sixteen kindergartners' gender-related views of careers.
\newblock \emph{The Elementary School Journal} 95(1): 95--103.

\bibitem[{Suresh and Guttag(2019)}]{suresh2019framework}
Suresh, H.; and Guttag, J.~V. 2019.
\newblock A Framework for Understanding Unintended Consequences of Machine
  Learning.

\bibitem[{Vorvoreanu et~al.(2019)Vorvoreanu, Zhang, Huang, Hilderbrand,
  Steine-Hanson, and Burnett}]{vorvoreanu2019gender}
Vorvoreanu, M.; Zhang, L.; Huang, Y.-H.; Hilderbrand, C.; Steine-Hanson, Z.;
  and Burnett, M. 2019.
\newblock From gender biases to gender-inclusive design: An empirical
  investigation.
\newblock In \emph{Proceedings of the 2019 CHI Conference on Human Factors in
  Computing Systems}, 1--14.

\bibitem[{Wang, Harper, and Zhu(2020)}]{wang2020factors}
Wang, R.; Harper, F.~M.; and Zhu, H. 2020.
\newblock Factors Influencing Perceived Fairness in Algorithmic
  Decision-Making: Algorithm Outcomes, Development Procedures, and Individual
  Differences.
\newblock In \emph{Proceedings of the 2020 CHI Conference on Human Factors in
  Computing Systems}, 1--14.

\bibitem[{White and White(2006)}]{white2006}
White, M.; and White, G. 2006.
\newblock Implicit and Explicit Occupational Gender Stereotypes.
\newblock \emph{Sex Roles} 55: 259--266.

\bibitem[{Woodruff et~al.(2018)Woodruff, Fox, Rousso-Schindler, and
  Warshaw}]{woodruff2018qualitative}
Woodruff, A.; Fox, S.~E.; Rousso-Schindler, S.; and Warshaw, J. 2018.
\newblock A qualitative exploration of perceptions of algorithmic fairness.
\newblock In \emph{Proceedings of the 2018 chi conference on human factors in
  computing systems}, 1--14.

\bibitem[{Woodworth et~al.(2017)Woodworth, Gunasekar, Ohannessian, and
  Srebro}]{woodworth2017learning}
Woodworth, B.; Gunasekar, S.; Ohannessian, M.~I.; and Srebro, N. 2017.
\newblock Learning non-discriminatory predictors.
\newblock \emph{arXiv preprint arXiv:1702.06081} .

\bibitem[{Yan et~al.(2020)Yan, Gu, Lin, and Rzeszotarski}]{yan2020silva}
Yan, J.~N.; Gu, Z.; Lin, H.; and Rzeszotarski, J.~M. 2020.
\newblock Silva: Interactively Assessing Machine Learning Fairness Using
  Causality.
\newblock In \emph{Proceedings of the 2020 CHI Conference on Human Factors in
  Computing Systems}, 1--13.

\bibitem[{Yao and Huang(2017)}]{Yao2017FairCF}
Yao, S.; and Huang, B. 2017.
\newblock Beyond Parity: Fairness Objectives for Collaborative Filtering.
\newblock In \emph{Proceedings of the 31st International Conference on Neural
  Information Processing Systems}, NIPS’17, 2925–2934. Red Hook, NY, USA:
  Curran Associates Inc.
\newblock ISBN 9781510860964.

\bibitem[{Youyou, Kosinski, and Stillwell(2015)}]{youyou2015computer}
Youyou, W.; Kosinski, M.; and Stillwell, D. 2015.
\newblock Computer-based personality judgments are more accurate than those
  made by humans.
\newblock \emph{Proceedings of the National Academy of Sciences} 112(4):
  1036--1040.

\bibitem[{Zafar et~al.(2017)Zafar, Valera, Gomez~Rodriguez, and
  Gummadi}]{zafar2017fairness}
Zafar, M.~B.; Valera, I.; Gomez~Rodriguez, M.; and Gummadi, K.~P. 2017.
\newblock Fairness beyond disparate treatment \& disparate impact: Learning
  classification without disparate mistreatment.
\newblock In \emph{Proceedings of the 26th international conference on world
  wide web}, 1171--1180.

\bibitem[{Zhang, Lemoine, and Mitchell(2018)}]{zhang2018mitigating}
Zhang, B.~H.; Lemoine, B.; and Mitchell, M. 2018.
\newblock Mitigating unwanted biases with adversarial learning.
\newblock In \emph{Proceedings of the 2018 AAAI/ACM Conference on AI, Ethics,
  and Society}, 335--340.

\end{thebibliography}
\end{small}
\end{quote}

\end{document}